# Tailoring the local density of states of non-radiative field at the surface of nanolayered materials

**Philippe Ben-Abdallah[1], Karl Joulain[2], Jérémie Drevillon[2] and Gilberto Domingues[1]**

1) Laboratoire de Thermocinétique, CNRS UMR 6607, Ecole Polytechnique de l'Université de Nantes, 44 306 Nantes cedex 03, France.

2) Laboratoire d'Études Thermiques (LET)-ENSMA, 1 Avenue Clément Ader, BP 40109, 86961 Futuroscope Chasseneuil Cedex, France.

**Abstract**

The ability to artificially grow in a controllable manner at nanoscale, from modern deposition techniques, complex structural configurations made with metallic, polar and semiconductors materials raises today the issue of the 'best' achievable inner structure to tailor the near-field properties of a nanostructured medium. In the present work we make a step towards the rational design of these materials by reporting numerical experimentations demonstrating the possibility of identifying structural configurations of layered metallodielectric media specifically designed to control the electromagnetic density of states in the close vicinity of their surface. These results could find broad applications in near-field technologies.

[1] Electronic mail : pba@univ-nantes.fr

[2] Electronic mail : karl.joulain@univ-poitiers.fr

One of the most important challenges in physics for the next decades is probably the conception of high performance materials for developing breakthrough energy technologies. Engineered materials at the scale of correlation lengths of energy carriers open compelling prospects to get innovating properties for energy transport that we don't generally find as it in nature. Among these properties, the electromagnetic local density of states (LDOS) which describes all optical modes accessible for photons in a specific region of space, conditions numerous quantum optical mechanisms such as the decaying rate of spontaneous emission of radiating dipoles or the radiative lifetime of molecules close to a surface[1]. It also enables us to deduce many statistical and thermodynamics properties of material such as its partition function, its heat capacity or its free energy. Since the pionner works of Purcell[2] on the LDOS enhancement and more recently of John and Yablonovitch[3,4] on its inhibition, we know that the DOS in a given region of space is strongly conditionned by its close environment. The ability to artificially grow in a controllable manner, from modern deposition techniques, complex structural configurations raises today the issue of the LDOS tailoring inside and around a medium. In particular, the occurrence and the increasing of local density of evanescent modes tends to dramatically affect the LDOS at the immediate proximity of the structure. In the present Letter we propose a rational method for the design of nanolayered metallodielectric near-field sources in a beforehand defined spectral region. Our approach combine the fluctuationnal electrodynamics theory[5] to calculate the LDOS at an arbitrary distance from the surface of any multilayered structure (direct problem) to an evolutionary algorithm[6] use to select in the set of all possible structural configurations, the best inner structure to achieve a near-field source with a target density of states (inverse problem).

As starting point, let us describe the calculation of the electromagnetic LDOS of a composite structure built by superposing plane layers made of N distinct nonmagnetic materials formed from M unit layers of the same thickness. We also assume that these structures are immersed in lossless dielectrics. Accordingly, the structure can be associated to a M-vector $\mathbf{S} = \{i_1,...,i_M\}$ where each $i_j$ identifies one of N basic materials. Because of the presence of thermal fluctuations inside the film, its local charges randomly oscillate and give rise to electric and magnetic fields **E** and **H** outside the structure. According to the linearity of Maxwell's equations, these fields are related to local fluctuating electric currents **j** through the following relations

$$\mathbf{E}(\mathbf{r},\omega) = -i\omega\mu_0 \int_V d\mathbf{r}' \overline{\overline{\mathbf{G}}}_{EE}(\mathbf{r}',\mathbf{r},\omega) \cdot \mathbf{j}(\mathbf{r}',\omega),\qquad(1\text{-a})$$

$$\mathbf{H}(\mathbf{r},\omega) = \int_V d\mathbf{r}' \nabla_{r_2} \times \overline{\overline{\mathbf{G}}}_{EE}(\mathbf{r}',\mathbf{r},\omega) \cdot \mathbf{j}(\mathbf{r}',\omega),\qquad(1\text{-b})$$

where $\overline{\overline{\mathbf{G}}}_{EE}(\mathbf{r}',\mathbf{r},\omega)$ represents the dyadic electric Green tensor of multilayered system[7] at point $\mathbf{r}$ in the surrounding medium from a point source $\mathbf{r}'$ and V denotes the total volume of the emitting body. At equilibrium, the LDOS $\rho(\mathbf{r},\omega)$ of electromagnetic field is related to the local density of energy $u(\mathbf{r},\omega)$ by the relation $\rho(\mathbf{r},\omega) = u(\mathbf{r},\omega)\Theta^{-1}(\omega,T)$ where $\Theta(\omega,T) \equiv \hbar\omega/[\exp(\hbar\omega/k_BT)-1]$ is the mean energy of a Planck oscillator at equilibrium temperature $T$. As for the density of energy, it is readily calculated from the spatial correlations of electric and magnetic fields

$$u(\mathbf{r},\omega) = \varepsilon_0 \langle \mathbf{E}(\mathbf{r},\omega) \cdot \mathbf{E}^*(\mathbf{r},\omega) \rangle + \mu_0 \langle \mathbf{H}(\mathbf{r},\omega) \cdot \mathbf{H}^*(\mathbf{r},\omega) \rangle,\qquad(2)$$

where the bracket denotes the statistic average over the ensemble of all realizations of random currents and where $\varepsilon_0$ and $\mu_0$ stand for the vacuum permittivity and permeability, respectively. Using relations (1) and the fluctuation dissipation theorem

$$\langle j_n(\mathbf{r},\omega) \cdot j_m^*(\mathbf{r}',\omega) \rangle = \frac{\omega\varepsilon_0\varepsilon''(\omega)\Theta(\omega,T)}{\pi}\delta_{nm}\delta(\mathbf{r}-\mathbf{r}')\qquad(3)$$

which relates the spatial correlation of fluctuating currents to the mean energy $\Theta(\omega,T)$, it is possible to derive[8] using an optical reciprocity theorem[9] the following general form for the LDOS at a distance $z$ from the surface

$$\rho(z,\omega) = \rho_{vacuum} \int_0^{\omega/c} \frac{k_{//}dk_{//}}{k_0|\gamma|} \frac{2+(k_{//}/k_0)^2\{Re[r^s \exp(2i\gamma z)] + Re[r^p \exp(2i\gamma z)]\}}{2}$$
$$+ \rho_{vacuum} \int_{\omega/c}^{\infty} \frac{4k_{//}^3 dk_{//}}{k_0^3|\gamma|} \frac{Im[r^s] + Im[r^p]}{2} \exp[2\,Im(\gamma)z]\qquad(4)$$

where $\rho_{vacuum} = \frac{\omega^2}{\pi c^3}$ is the familiar free-space electromagnetic LDOS. In this expression, the first integral represents the contribution of propagative waves while the second one denotes that of non-radiative modes. Here, $r^{s,p} = r^{s,p}[\mathbf{S}]$ represents the reflectivity of multilayered system respectively for shear (s) and parallel (p) waves, $\gamma = \sqrt{\varepsilon_{sur} k_0^2 - \mathbf{k}_\parallel^2}$ is the normal component of wave vector in the surrounding medium and $\mathbf{k}_\parallel$ its parallel component. Since the reflectivity factors are the only functions which depend on the inner structure of medium we see that expression (4) establishes a straightforward relation between $\mathbf{S}$ and the LDOS in the surrounding medium.

To identify the "best structure" which leads to a LDOS as close as possible than a target LDOS $\rho_{targ}(z,\omega)$ we have to inverse the strongly non-linear integral equation (4) that is, in others words, to find an optimal vector $\hat{\mathbf{S}}$ which satisfies

$$J \equiv \left\| \rho[\hat{\mathbf{S}}] - \rho_{targ} \right\| \to min. \tag{5}$$

Before solving this problem let us first remark that the total number of all possible configurations achievable with $M$ layers of $N$ distinct materials is $N^M$. Hence, if we wish, for instance, to design a binary structure made with 32 unit layers then we have $2^{32} \sim 10^{10}$ possible configurations. This number even reaches about $10^{30}$ (one million times the Avogadro number!) with 100 layers. Such a large space of configurations offers immense possibilities to sculpt the LDOS inside and around the structure. However, to explore efficiently this vast space requires the use of a rational searching method. To this end we apply a genetic algorithm[6] (GA)-based optimization process which is a global optimisation algorithm well adapted to solve nonlinear inverse problem. The GA use the basic principles of natural selection rules like in the Darwin theory of evolution. At each step of this optimization process, the discrepancy between the objective function $\rho_{targ}$ and the calculated LDOS is formally measured by the fitness function J. Some mutations are also randomly introduced at each generation with an occurrence probabilbity p in order to improve the GA convergence. This mutation parameter is adjusted each K generations using the probability increment $\Delta p = \pm \xi$ .

Now we apply this procedure to design binary structures which possess a LDOS as close as possible than a given target LDOS. In this work, these structures are designed with 32 superposed unit layers 5 nm thick each and composed either by aluminium or a lossless material. Moreover, they are optimized to operate in the near-UV range $I = [0.65\omega_p ; 0.85\omega_p]$ close to the plasma frequency $\omega_p = 17.47 \times 10^{15}$ rad.s$^{-1}$ of aluminium. Because of their size, each Al layers which compose the composite structures, support two non-degenerated surface plasmons (SP), the so called symmetric (low frequency) and antisymmetric (high frequency) surface modes[10]. These modes are spatially localized on each layer, respectively at lower and higher frequency than the bulk surface plasmon frequency $\omega_p / \sqrt{2}$ . Therefore, a complex composite structure is generally made with a wide population of different SPs and hybridization between these modes inside the structure allows to modify the LDOS in the surrounding medium.

In the near-UV range, the Thomas-Fermi screening length $v_F / \omega_p$ and the Fermi wavelength $v_F / \omega_c$ , defined with $v_F = 2.03 \times 10^6 m.s^{-1}$ , the Al Fermi velocity and $\omega_c = 7.596 \times 10^{13}$ rad.s$^{-1}$ the electron collision frequency, are both of the order of nanometer. Therefore, we can neglect the non-local effect and describe the Al dielectric permittivity by the free electron Drude model[11] $\varepsilon_{Al} = 1 - \frac{\omega_p^2}{\omega(\omega - i\omega_c)}$ .

For the seek of simplicity, the losseless material we used to compose these structures is assumed non-dispersive and supposed to have a dielectric constant greater than one to naturally enhance the spontaneous emission rate[12]. Otherwise, in a complex lattice of localized modes, the LDOS is the result of the interaction of each aluminum layer surface modes. These modes are distributed around the bulk aluminum SP frequency so that the interaction between these close modes gives also, in the weak interaction approximation (the attenuation length $\delta = 1/2 Im[\sqrt{\varepsilon k_0^2 - k_{//}^2}]$ of symmetric and antisymmetric modes in the lossless material is of the order of nanometer), a distribution of modes around the surface plasmon frequency. Thus, we have chosen as target LDOS a Lorentzian function[13] of the form $\rho_{targ}(\omega) = \rho_{max} \frac{\Delta\omega^2}{(\omega - \omega^*)^2 + \Delta\omega^2}$ [$\rho_{max}$ defines the magnitude of the Lorentzian, $\Delta\omega$ its width at half height and $\omega^*$ the localization of its transition frequency] which is centred either at $\omega^* = \omega_p/\sqrt{2}$ (Fig. 1-a), the frequency of Al surface plasmon (surrounded by vacuum), or shifted at $\omega^* + \delta\omega$ (Fig.1-b). The maximum density of this target is chosen equal to $\rho_{max} = 3 \times 10^9$ that is approximately 10 times the maximum value of the LDOS measured at the same distance from a single Al layer 160 nm thick. After 200 generations of evolution we observe (Fig.1), using a fitness function defined as $J = [\int_I (\rho(\omega;z) - \rho_{targ}(\omega))^2 d\omega]^{1/2}$, that the targets and the optimized LDOS differ by less than one order of magnitude everywhere on I, in both cases. Moreover, we see on the same figure, by displaying the LDOS at 10 nm from the surface of the half-structured composite medium, made with the whole optimized structure where each Al layer in the second half has been replaced by a layer of lossless material, that it is mainly the SPs interactions in the first half which condition the LDOS shape. We see also (Fig.1-b) that to shift the LDOS peak toward higher frequencies than the bulk SP frequency thinner Al layers are required in the first half of the structure (Fig. 2). In this case, the spectral shift between the symmetric and antisymmetric SPs supported by each Al layers increases so that the high frequency mode moves upward well beyond the bulk SP frequency [notice that there is also a broader secondary peak at low frequency in the LDOS (not plotted) because of the presence of symmetric SPs around $\omega = 0.5\omega_p$]. Moving furthermore the LDOS peak toward the high frequencies would necessitate even thinner basic layers. In this case, the nonlocal effects in the electromagnetic response of Al layers should be taken into account.

Finally the sensitivity of design process to changes in the optical properties of basic materials or to a perturbation of the inner structure has been investigated. To this end, we have either perturbated uniformly the Al dielectric permittivity with a 10% deviation random function $\theta(\omega)$, changed the value of $\varepsilon$ by modifying its magnitude or by introducing residual losses and finally we have changed the inner structure, by randomly modifying the thickness of unit layers in the optimized structure with a uniform random deviate of 20%. The results display on (Fig. 3), show a quite good robustness of optimization process for weak perturbations.

To conclude we think that, nanolayered materials offer immense possibilities for the control of near field properties from UV to IR ranges. Indeed, the presence of localized resonances modes such as surface plasmons or phonon-polaritons which are able, by changing the width and the positions of layers, to interact together and to strongly absorb light in different region of spectrum give rise to unique optical properties and allow an efficient control of the LDOS. The design of this LDOS in the vicinity of nanostructured objects is very promising to improve the performance of numerous near-field technologies such as the near-field thermophotovoltaïc conversion[14], the plasmon assisted nanophotolitography[15] or the near-field spectroscopy[16]. Since the LDOS is a fundamental quantity our approach could also be used to sculpt the thermodynamic properties of materials. Finally, a similar demarche could be followed for tailoring the Casimir force[17] and the non-contact friction forces[18] between nanostructured materials.

**Captions List**

Fig. 1: Local density of states of non-radiative modes at a distance z=10nm from the surface of an aluminium film 160 nm thick (green curve) and at the same distance from the surface of a multilayer medium composed by 32 unit layers 5 nm thick of aluminium or lossless material ($\varepsilon = 2.25$) and designed by GA to retrieve the Lorentzian LDOS plotted in red. The population size is equal to 10, the number of generation is 200 and the mutation probability is adjusted each K=$10$ generations with $\xi = 10^{-3}$. The dash-doted curves represent the LDOS at 10 mn from the surface of the half-structured medium. For the first target (a), the maximum of the LDOS is centered at $\omega^* = \omega_p / \sqrt{2}$. It is shifted toward $\omega^* + \delta\omega$ with $\delta\omega = 0.02\omega_p$ for the second target (b).

Fig. 2 : Minimum of the fitness function vs. the number of generation and optimized structure after 200 generations for two different target LDOS. The red lames denote aluminum layers and the blue ones symbolize the lossless material layers ($\varepsilon = 2.25$). Each unit layer is 5 nm thick and the total thickness of the optimized structure is 160 nm.

Fig. 3: LDOS sensitivity (corresponding to the case illustrated on Fig.1-b) to the shape of the metal (Al) dielectric permittivity, to residual losses in the non-metallic material and to the multilayer structure. The inset shows the LDOS sensitivity to the magnitude of the dielectric constant $\varepsilon$.

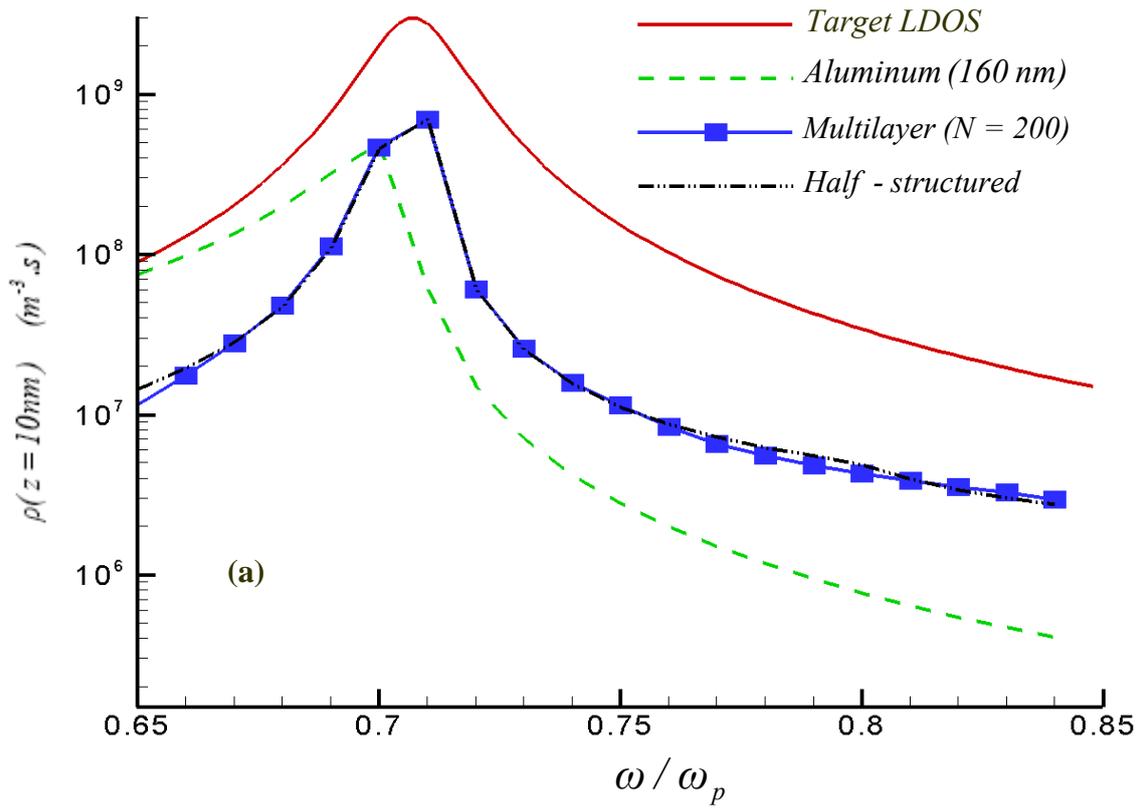

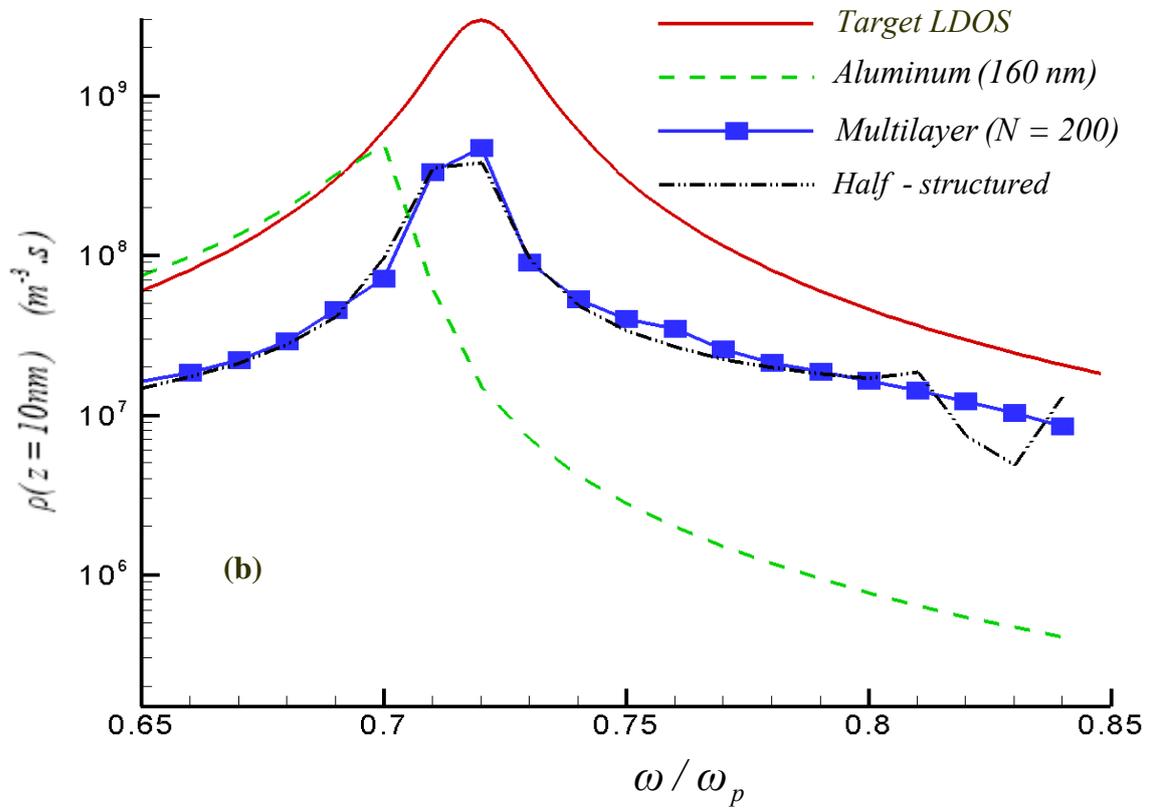

**Figure 1**

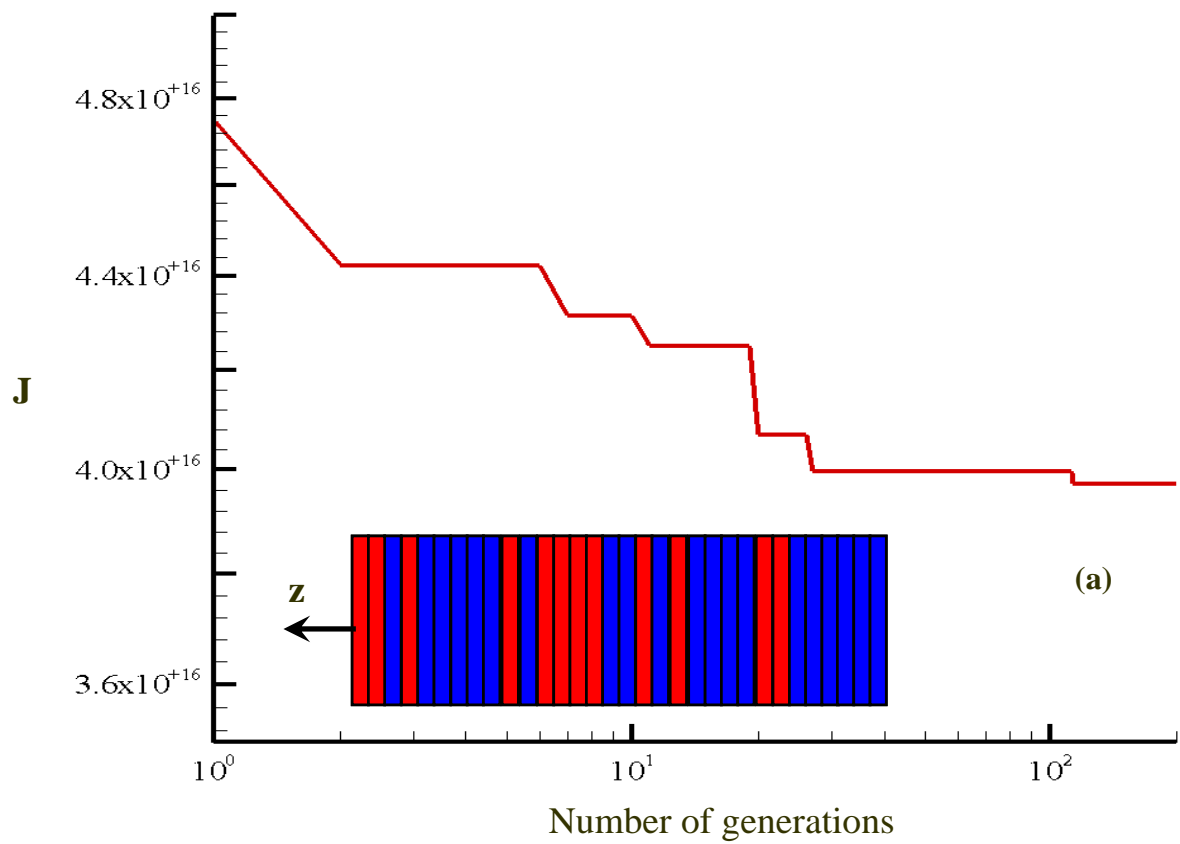

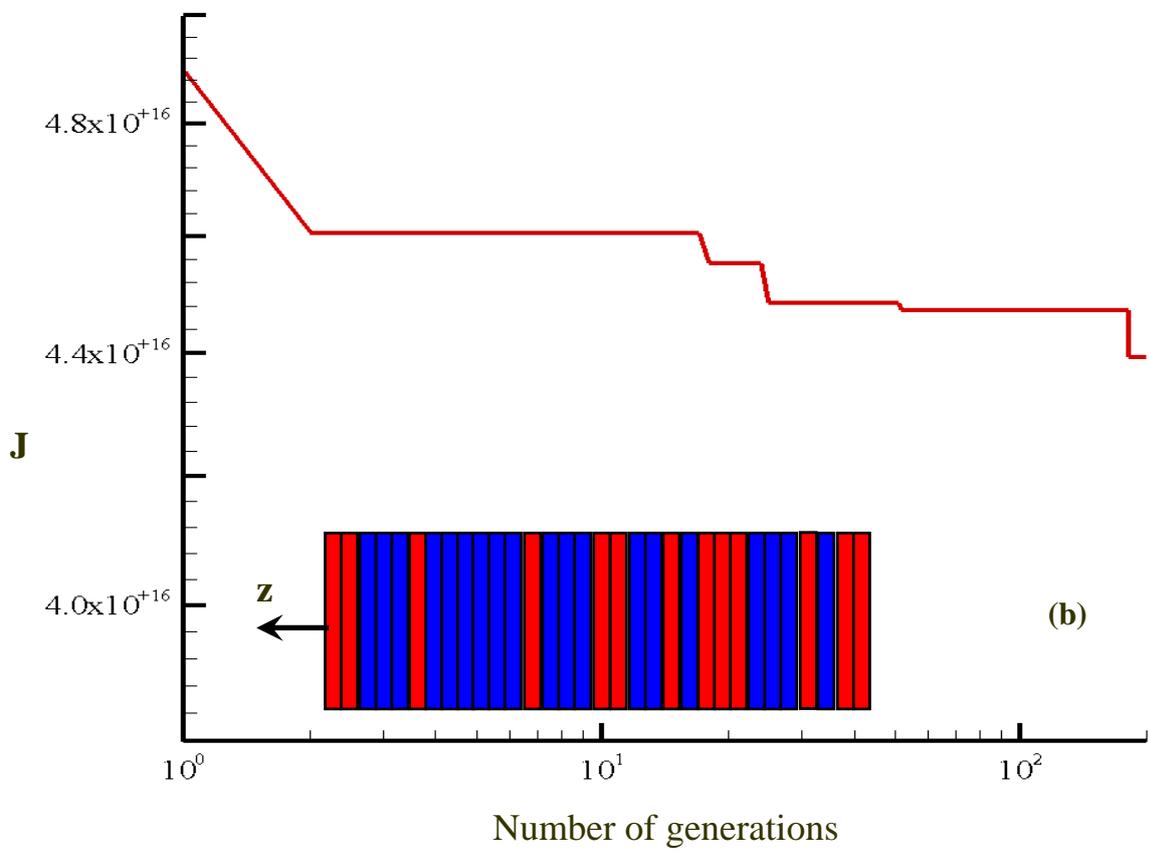

**Figure 2**

z

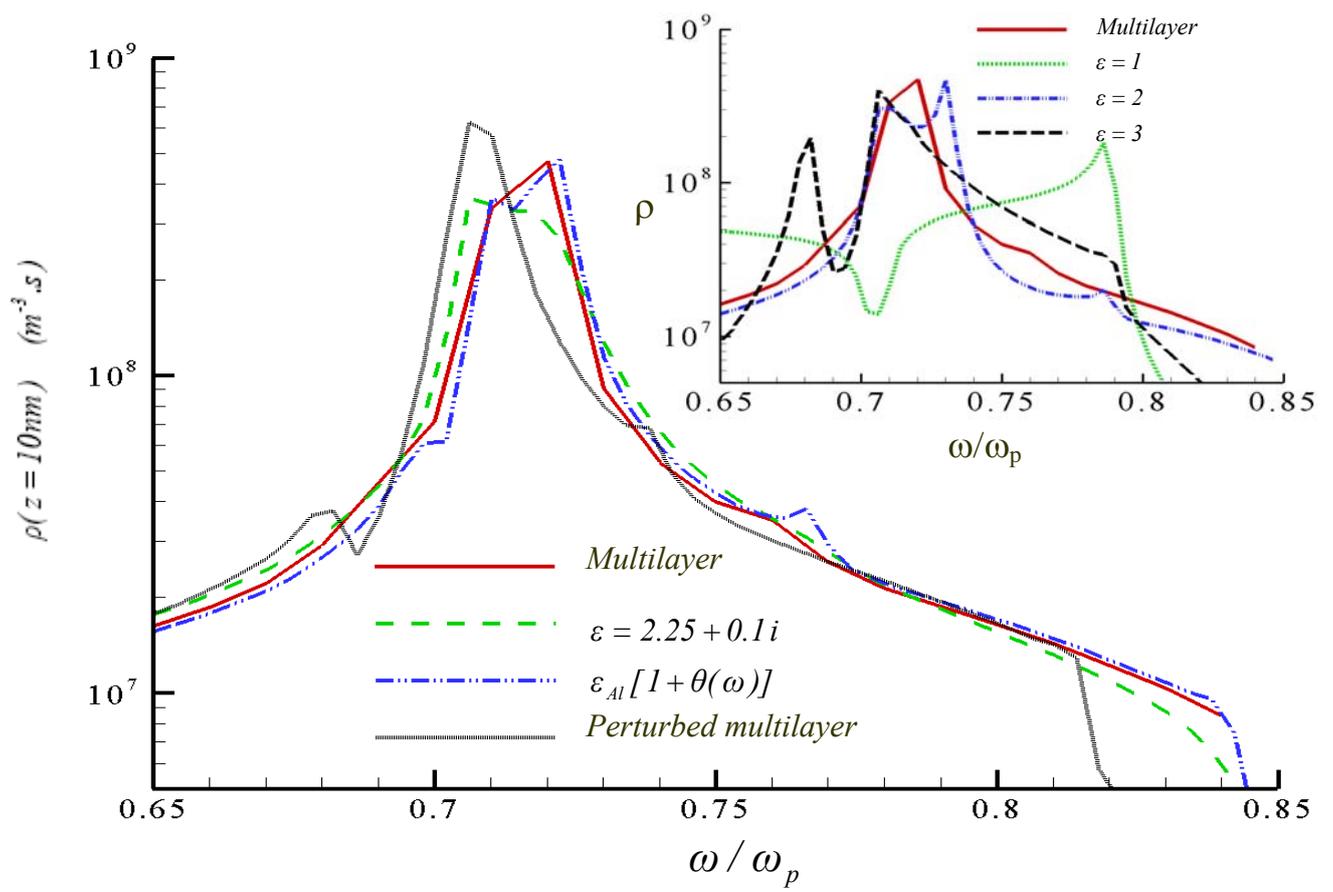

**Figure 3**